\documentclass[12pt,a4paper]{article}

\usepackage{graphicx}
\usepackage{amsmath}
\usepackage{amsfonts}
\usepackage{amssymb}
\def\simleq{\; \raise0.3ex\hbox{$<$\kern-0.75em \raise-1.1ex\hbox{$\sim$}}\; }
\def\simgeq{\; \raise0.3ex\hbox{$>$\kern-0.75em \raise-1.1ex\hbox{$\sim$}}\; }
\newcommand{\eV}{{\rm eV}}

\newcommand{\GeV}{{\rm GeV}}
\newcommand{\TeV}{{\rm TeV}}
\newcommand{\PeV}{{\rm PeV}}

\newcommand{\kpc}{{\rm kpc}}
\newcommand{\pc}{{\rm pc}}
\newcommand{\cm}{{\rm cm}}

\newcommand{\muG}{\mu{\rm G}}

\newcommand{\s}{{\rm s}}

\newcommand{\anu}{\overline{\nu}}

\begin{document}
 \title{TeV Neutrinos from SuperNova Remnants embedded in Giant Molecular Clouds}
 \author{Vincenzo Cavasinni$^{1,2~\dag}$,  \ Dario Grasso$^{2,3~\ddag}$, \ Luca Maccione$^{2,4~*}$}
\maketitle
\noindent
$^1$ Dip. di Fisica ``E.~Fermi", Universit\`a di Pisa, Largo B.~Pontecorvo, 3,  Pisa\\
$^2$ I.N.F.N. Sezione di Pisa\\
$^3${ Scuola Normale Superiore, P.zza Dei Cavalieri, 7, I-56126 Pisa}\\
$^4$ S.I.S.S.A., Via Beirut, 2- I-34014, Trieste
\vskip0.3cm \noindent
$^\dag$ Vincenzo.Cavasinni@pi.infn.it; 
$^\ddag$ d.grasso@sns.it; \ $^*$ luca.maccione@pi.infn.it
\noindent
\begin{abstract}
The recent detection of  $\gamma$-rays with energy up to 10 TeV from dense regions 
surrounding  some Supernova Remnants  (SNR) provides strong, though still not conclusive, 
evidence that the nucleonic component of galactic Cosmic Rays 
is accelerated in the supernova outflows.  Neutrino telescopes could  further support
the validity of  such scenario by detecting neutrinos  coming from the same regions. 
We re-evaluate the TeV range neutrino-photon flux ratio to be expected from pion decay, finding
small differences respect to previous derivations. 
 We apply our results and the recent HESS measurements of the very high energy
  $\gamma$-ray flux from the molecular cloud complex in the Galactic Centre,  
  to estimate the neutrino flux from  that region discussing the detectability perspectives
  for Mediterranean Neutrino Telescopes. 
We also discuss under which conditions  neutron decay may give rise to a  significant 
TeV antineutrino flux from a SNR  embedded in Molecular Cloud complexes.  
 \end{abstract}

\section{Introduction}

The existence of a    correlation between the position of supernova remnants (SNRs)
 and that of giant molecular clouds (GMCs) is supported by theoretical arguments and
observational evidences.
Infrared observations confirmed the theoretical expectation that stars,
especially massive ones, form mainly in the womb of GMs.
Having a relatively short timelife, massive stars can complete their
evolution and explode within the parent GMC before this is
swept-out by stellar winds. SN explosions within, or in the nearby of, GMCs
may trigger the formation of new stars explaining the bursting star formation activity
observed in several dense regions of the Galaxy, e.g. in the Orion giant cloud,
and in starburst galaxies.
  
From the point of view of high energy astrophysics,  SNRs interacting with MCs
 are very interesting systems. Since SN explosions are expected to be the main
source of the bulk of cosmic rays (CRs) in the InterStellar Medium (ISM) \cite{Ginzburg},
a MC which is embedding, or just lay in the nearby of, an active SNR, should be crossed
by a flux of relativistic particles which is significantly larger than in the rest of the Galaxy.
Furthermore, due to their relatively high gas density ($n_H \sim 10^2 \div 10^6~\cm^{-3}$),
GMCs should act as efficient astrophysical beam-dumps converting a relevant fraction
of the primary particle energy into secondary $\gamma$-ray and neutrino emissions
(see e.g. \cite{Montmerle,Aharonian96}). In the following we will generally call SNRs
interacting with MCs as {\it embedded SNRs}, also in those cases in which the SNR is not completely
surrounded by the molecular gas.

Evidences of a correlation between $\gamma$-ray emission and dense molecular gas
have been found by several satellites and particularly by EGRET (see \cite{Egret}) and have been extensively studied by many authors (see e.g. \cite{Torres:2002af}).
 Those observations, however, may generally
be explained without invoking any local excess in the CR energy density respect to the value observed
on the Earth. Only recently, a new generation of atmospheric Cherenkov telescopes, especially
  HESS, were able to find the smoking-gun of the interaction of CRs accelerated by an
  active SNR with a close MC. The most interesting HESS observations in this respect
 are those of RXJ1713.7-3946 \cite{RXJ1713} and the very recent measurements of the Galactic
 Centre (GC) ridge emission \cite{Hess2006}.

Several arguments suggest that the origin of the very high energy (HE) radiation coming from these
sources may be hadronic.  One of the most solid among these arguments is based on the
 observation of strong (up to several milliGauss)  magnetic fields in MCs \cite{Heiles:2005hr} (see  \cite{YZ96} for  the GC region). 
 Such strong field   should give rise to dramatic electron synchrotron energy losses above the TeV suppressing a possible  HE Inverse Compton contribution to  the total
 $\gamma$-ray emission from those regions (see e.g. \cite{Hess2006}).
Although the previous argument is rather convincing, a more direct evidence that HE
$\gamma$-rays from embedded SNRs are produced by $\pi^0$ decay would be extremely precious to confirm the validity of the entire scenario.
Such an evidence may come from the detection of neutrinos coming from the same sources.

Several Neutrino telescopes are already operating, or under construction, around the world 
(see e.g. \cite{Montaruli:2005fa}). 
Due to their large detection volume and observation time, these instruments may have the
capability to detect at least few neutrinos from some SNRs, especially if they are 
embedded in a dense medium.
It is important to observe that the energy threshold above which Neutrino Telescopes may identify astrophysical neutrino sources ($\simgeq 100~ \GeV$) is approximatively
the same above which
atmospheric Cherenkov telescopes can detect photons. This coincidence offers the interesting possibility to directly test if HE photons and neutrinos have the same hadronic 
origin confirming the half-century  pending hypothesis that HECRs are accelerated by SNRs.
Having that purpose in mind, in this paper we estimate the expected neutrino/photon flux ratio in the TeV energy range from embedded SNRs.

Our computation of the $\nu/\gamma$ ratio from pion decay is similar to that followed by the authors of \cite{Costantini:2004ap} for RXJ1713-3946 
(see also \cite{Costantini:2005vh}) and confirm their main results,
 a part small differences which we discuss in Sec.\ref{sec:nuPion}.

In Sec.\ref{antinu} we discuss the production of TeV antineutrinos by neutron decay in 
GMC complexes. In \cite{Anchordoqui04} it was proposed that this process may
 give rise to a detectable antineutrino flux from the Cyg OB2 association and from the GC.
 That analysis was based on Ultra High Energy CR anisotropy measurements which, 
however, have a very low statistical significance or, in the case of the GC, have not been 
confirmed by other experiments.
 We will show that unless the primary particle spectrum is considerably different from
 that expected from ordinary shock acceleration, an observable contribution of neutron 
decay to the total neutrino flux can hardly be expected. 
Finally, in Sec.\ref{sec:nuGC} we apply the results of Sec.\ref{sec:nuPion} to estimate the muon
neutrino flux from the MC complex in the GC on the basis of the new HESS measurements
\cite{Hess2006}.
 
\section{Photon emissivity by $\pi_0$ decay}
 
 The process under consideration is $pp \rightarrow NN\,+\,{\rm n}_{\pi}\pi^{\pm,0}$
 where ${\rm n}_{\pi}$ is the pion multiplicity and $N$ is either a proton or a neutron.
The general expression of the photon emissivity is
\begin{equation}
Q_{\pi^0}(E_ {\pi^0}) = c \; n_H
\int_{E_p^{min}(E_{\pi^0})}^{E_p^{max}}dE_p ~n_p(E_p)
\frac{d\sigma}{dE_{\pi^0}}(E_p, E_{\pi^0})~,
\end{equation}
where we assumed that hydrogen, having density $n_H$, is the main proton target.
When the proton energy is much larger than the proton mass,
i.e. $E_p \simgeq {\rm several}~\GeV$, the differential pion production
cross section is well approximated by a scaling expression
\begin{equation}
\label{ph_scaling}
{d\sigma (E_p, E_{\pi^0})\over dE_{\pi^0}} = {\sigma_0\over
E_{\pi^0}} f_{\pi^0} (x)\;,
\end{equation}
where $x \equiv E_{\pi^0}/E_p$ and $\sigma_0 \simeq 3 \times 10^{-26}~\cm^{-2}$. 
Using the numerical package \textsc{pythia} \cite{Sjostrand:2003wg} we verified 
that the expression of the scaling function adopted by the authors of \cite{Blasi:1999aj},
 $f _{\pi^0} (x) = 0.67(1-x)^{3.5} + 0.5e^{-18x}$, 
provides a good fit of simulated $pp$ scattering data up to $500~\TeV$. 

In all cases in which the proton spectrum is well approximated by a power spectrum
\begin{equation}
\label{p_spectrum}
n_p(E_p) = n_0 \left(\frac{E_p}{E_0}\right)^{-\alpha}\;,
\end{equation}
the $\pi^0$ emissivity becomes
\begin{equation}
\label{q_gamma}
Q_ {\pi^0}(E_ {\pi^0}) \simeq
n_p(E_\pi^0) \sigma_0~ c \; n_H Y(\alpha)~,
\end{equation}
where
\begin{equation}
\label{gamma_yield}
Y(\alpha) = \int_0^1 dx x^{\alpha-2} f_{\pi^0}(x, \alpha)
\end{equation}

Hence the photon emissivity is
\begin{equation}
\label{q_gamma}
\begin{split}
Q_\gamma(E_\gamma) & = \int_{E_ {\pi^0}^{\rm min}(E_\gamma)}
 dE_ {\pi^0} \frac{2}{E_ {\pi^0}} Q_ {\pi^0}(E_ {\pi^0})
\\
& \simeq \frac{2}{\alpha}
   \sigma_0~ c \; n_H Y(\alpha)~n_p(E_\gamma)~,
   \end{split}
\end{equation}
where we used $\displaystyle E_ {\pi^0}^{\rm min}(E_\gamma) =
E_\gamma \left( 1 + \frac{m_{\pi^0}^2}{4 E_\gamma^2}\right)$.
  
 It is also useful to write the mean value of the ratio between photon energy
and that of the parent proton ({\it photon elasticity}), which is
\begin{equation}
\label{ ph_elasticity}
\begin{split}
\langle E_\gamma/E_p \rangle & =
\frac{1}{2} \int_{-\pi}^{\pi} d\cos \theta ~ (1 - \cos \theta) ~\int_0^1 dx
f_{\pi^0}(x)\\
& = \int_0^1 ~dx ~f_{\pi^0}(x) = 0.18
\end{split}
\end{equation}
 where $\theta$ is the angle of the photon momentum in the lab frame respect to that of  the primary proton.
We conclude this section by observing that by writing the proton spectrum  in the form Eq. (\ref{p_spectrum}), we implicitly assumed that the ultra-violet cut-off is well above  
the maximal energy relevant for $\gamma$-ray observations. This is well motivated 
by the fact that the energy spectra observed by HESS for  RXJ1713.7-3946 
\cite{RXJ1713} and the GC \cite{Hess2006}, which we assume to be representative 
of all embedded SNRs,  are steady power laws up to $\sim 10~\TeV$.   

\section{Muon and Electron Neutrinos from charged pion decay}\label{sec:nuPion}

\begin{figure}
   \centering
   \includegraphics[scale=0.5]{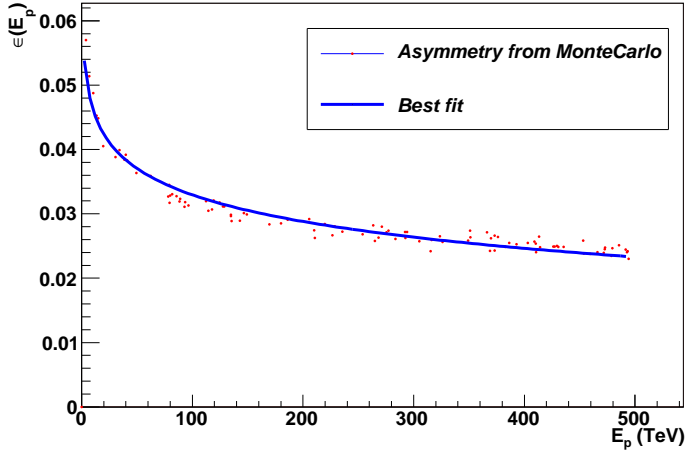}
   \caption{Energy dependence and best fit of $\epsilon$, as provided by \textsc{pythia}.}
   \label{fig:epsilon}
\end{figure}

Inelastic $pp$ collisions lead to roughly the same number of $\pi^0$'s,
$\pi^+$'s and $\pi^-$'s . If, as we are assuming here, $\pi^0$ decay is the
dominant source of high energy photons, a comparable emission of $\nu_\mu$ and
 $\nu_e$ has to be expected.
The emissivity of muon neutrinos and antineutrinos
produced by direct $\pi^+$ and $\pi^-$decay is
\begin{equation}
\label{nu_pi_Q}
Q^\pi_{\nu_\mu}(E_{\nu_\mu}) \simeq \frac{m_\pi^2}{m_\pi^2-m_\mu^2}\int_{E_\pi^{min}(E_\nu)}^\infty
\frac{dE_\pi}{E_\pi}Q_{\pi^\pm}(E_\pi)\;,
\end{equation}
where $Q_{\pi^\pm}(E_\pi) = Q_{\pi^0}(E_\pi)\times (1\pm\epsilon)$, $\epsilon$ being the asymmetry between the positive and negative pions effectively produced in $pp$ collisions:
\begin{equation}
\epsilon \equiv \frac{N_{\pi^+}-N_{\pi^-}}{N_{\pi^+}+N_{\pi^-}}\;.
\end{equation}
For $\langle E_\nu \rangle \gg m_\pi$,
\begin{equation}
E_\pi^{\rm min}(E_\nu) =
\frac{m_\pi^2}{m_\pi^2-m_\mu^2}E_\nu +
\frac{m_\pi^4}{m_\pi^2-m_\mu^2} ~\frac{1}{4\,E_\nu}
\simeq \frac{1}{1 - r^2}~E_\nu\;,
\end{equation}
where $r \equiv m_\mu/m_\pi$.
 Since we don't know the details of the  interaction, the parameter $\epsilon$ cannot be 
evaluated analytically  and we have to recourse to numerical simulations.
We performed a simulation of proton-proton interactions up to an incident energy of about 500 TeV with \textsc{pythia} \cite{Sjostrand:2003wg}. Our best fit for $\epsilon$ is
\begin{equation}
\epsilon(s) = 0.1-6\cdot10^{-3}\ln(s/s_{0})-0.15\cdot\sqrt{s_{0}/s}\;,
\end{equation}
with $s \equiv 2m_{p}^{2}+2m_{p}E_{p}$ and $s_{0} \simeq 1~\GeV^2$.
In Fig.\ref{fig:epsilon} we show the energy dependence of our $\epsilon$ as well as our best fit.
We can now write
\begin{equation}
\label{Q_nu_pi}
Q^\pi_{\nu_\mu}(E_\nu) \simeq \left(\frac{1}{1 - r^2}\right)^{1-\alpha}~\frac{1\pm\left\langle \epsilon\right\rangle }{2}Q_\gamma(E_\nu)~,
\end{equation}
where $\left\langle \epsilon\right\rangle$ represents the average of $\epsilon$ over the
 relevant energy range which, in our case, is $10~\GeV - 500~\TeV$.
The computation of the contribution of secondary muon decay
$\mu \rightarrow e \nu_\mu \nu_e$ to the neutrino emissivity
is slightly more involved.
The contribution of this process to the neutrino emissivity is
\begin{equation}
\label{Q_nu_mu}
Q_\nu^\pi (E_\nu) = \int_{E_\mu^{\rm min}(E_\nu)}^{\rm E_\mu^{\rm max}} ~ dE_\mu ~
Q_\mu^{\pi}(E_\mu) ~\frac{dP(E_\mu, E_\nu)}{dE_{\nu} }~,
\end{equation}
where from the kinematics it follows $E^{\rm min}_\mu \simeq E_\nu$ and $E^{\rm max}_\mu$ can be safely posed to $\infty$ if the muon spectrum is broad enough.
The muon emissivity is
\begin{equation}
\label{Q_mu}
Q^\pi_\mu(E_\mu) \simeq \frac{1}{1 - r^2} \int_{E_\mu}^{E_\mu/r^2}
\frac{dE_\pi}{E_\pi}Q_{\pi^\pm}(E_\pi) =
\frac{1}{1 - r^2} \left( 1 - r^{2\alpha}\right) \frac{1\pm\left\langle \epsilon\right\rangle }{2}Q_\gamma(E_\mu)\;.
\end{equation}
The neutrino emissivity can thus be estimated as done in \cite{book:Gaisser} by using the
spectrum-weighted momenta
\begin{equation}
\label{Q_mu_2}
\begin{split}
Q^\mu_\nu(E_\nu) \simeq ~& \frac{1\pm\left\langle \epsilon \right\rangle}{2}Q_\gamma(E_\nu)\frac{1-r^{2\alpha}}{\alpha(1-r^{2})}\times \\
& \left( \left\langle y^{\alpha-1}\right\rangle_0+\frac{1}{1-r^2}\left(1+r^2-\frac{2\alpha
 r^2}{\alpha-1}\frac{1-r^{2(\alpha-1)}}{1-r^{2 \alpha}}\right)\left\langle y^{\alpha-1}
  \right\rangle_1 \right)\;,
\end{split}
\end{equation}
where $\left\langle y^{\alpha-1}\right\rangle_0$ and $\left\langle y^{\alpha-1}\right\rangle_1$ are given in the same textbook for the different flavours.
The total neutrino emissivity by both positive and negative pion decay is given by the sum of (\ref{Q_nu_pi}) and of
 (\ref{Q_mu_2}).
  
We can account for  kaons contribution as well, provided that we can relate their flux to
 that of pions. Again, a simulation performed with the help of \textsc{pythia} allows us to determine that the mean ratio $K/\pi$ is about few percent. Thus, accounting for the well known branching ratio for the decay $K\rightarrow\mu\nu$, that is  63.5\%,  
 \footnote{We neglect here three-body decay which have only a few percent BR.}
 and keeping in mind that no significant dynamical differences arise between pions and kaons, 
we found the correction to the neutrino emissivity due to pion. This come out to be 
few percent.

Finally, using the expressions for the spectrum-weighted momenta and a mean asymmetry $\epsilon$ of about 5\% we find the values of the ratios $Q_{\nu_i}/Q_\gamma$ 
which we shown in Tab.\ref{tab:ratios} for several values of the spectral index $\alpha$.

\begin{table}[ht]
\begin{center}
\begin{tabular}{|c|c|c|c|c|}
\hline & $\alpha = 2.0$ & $ \alpha = 2.2$ & $\alpha = 2.4$ & $\alpha = 2.6$ \\
\hline $\nu_\mu$ & 0.52 & 0.46 & 0.40 & 0.36 \\
\hline $\anu_\mu$ & 0.52 & 0.46 & 0.40 & 0.36 \\
\hline $\nu_e$ & 0.28 & 0.25 & 0.22 & 0.20 \\
\hline $\anu_e$ & 0.26 & 0.23 & 0.21 & 0.18 \\
\hline
\end{tabular}
\end{center}
\caption{Ratio between neutrino and $\gamma$-ray emissivities
for several values of the primary proton spectral index $\alpha$. }
\label{tab:ratios}
\end{table}
  These values are close, though not coincident,   with those given in \cite{Costantini:2004ap}. 
The percentage difference ranges between 10\%  and 30\%, being maximal for  electron 
antineutrinos. Most likely, such a discrepancy is due  to a  value of the pion charge asymmetry
 that in \cite{Costantini:2004ap} is larger than the value adopted here.  
We think that the value of $\epsilon$ used in \cite{Costantini:2004ap} 
 is correct only at low energies ($\sim 100~ \GeV$) and that the authors 
 may have not taken into account the decreasing behaviour of $\epsilon$ at  larger energy.
 Indeed, from  Fig.1 we see that the scaling behaviour is reached only above several 
 hundred GeV.  
 Nevertheless, it should be noted that being the uncertainty 
 in the value of $\epsilon$ provided by  \textsc{pythia} of the order  of 20 \%, 
 the values of the neutrino-photon relative emissivities given in Tab. \ref{tab:ratios}
 are almost compatible with those given in  \cite{Costantini:2004ap} within errors.
  
  \subsection{The effect of Neutrino oscillation}
\label{subsec:NuOsc}
As well established by several experiments, neutrinos undergo flavor oscillations during their propagation in vacuum.
Since the typical SNRs  distance from the Earth is much larger than the neutrino
oscillation length around the  TeV,  the phase of oscillations is very large so that
we can safely deal with averaged vacuum oscillations. In this case the flavor oscillation probability can be written
\begin{equation}
P_{ll'} = \sum_{j}|U^2_{lj}|\cdot|U^2_{l'j}|\;,
\end{equation}
where $U_{ll'}$ is the unitary mixing matrix among the lepton flavours $l,l'=e,\mu,\tau$. After the propagation the neutrino flux is
\begin{equation}
\label{eq:osc}
F_i  = 
\sum_{j = e,\mu,\tau}P_{ij}F_j^0~,
\end{equation}
 $F_j^0$  being the expected neutrino flux in the absence of oscillations.  

Several experiments have been performed, and are now in progress, to determine
 the values of the mixing parameters. Adopting the standard parametrization \cite{Eidelman:2004wy} for the mixing matrix $U_{ij}$, we can describe our present knowledge as:
\begin{itemize}
\item $\theta_{12} \simeq \theta_{\odot} \simeq (33.9^{+2.4}_{-2.2})^{\circ}$ by solar and KamLAND data \cite{Aharmim:2005gt}.
\item $\theta_{23} \simeq \theta_{atm} \simeq 45^{\circ}\pm10^{\circ}$ by atmosferic data (see e.g. \cite{Costantini:2004ap}).
\item $\theta_{13} \simeq 0\pm10^{\circ}$ from CHOOZ data \cite{Apollonio:2002gd}.
\end{itemize}
The CP violating phase $\delta_{CP}$ is still unknown and  we assume here that it is
 negligible. Under this hypothesis the oscillation probability matrix $P_{ij}$ is symmetric and equal for both neutrinos and antineutrinos. We show it in Tab.\ref{tab:Osc}.

\begin{table}
\centering
\begin{tabular}{|c|c|c|c|}
\hline  & $\nu_e$ & $\nu_\mu$ & $\nu_\tau$ \\
\hline $\nu_e$ & 57\% & 21\% & 21\% \\
\hline $\nu_\mu$ & 21\% & 39\% & 39\% \\
\hline $\nu_\tau$ & 21\% & 39\% & 39\% \\
\hline
\end{tabular}
\caption{Oscillation probability in the average vacuum oscillation hypothesis, calculated with the mixing angles reported in the text.}
\label{tab:Osc}
\end{table}
Using Eq.(\ref{eq:osc}) we calculate the $\nu/\gamma$ ratios after propagation 
 which we shown in Tab. 3
\begin{table}
\begin{center}
\begin{tabular}{|c|c|c|c|c|}
\hline & $ \alpha = 2.0$ & $\alpha = 2.2$ & $\alpha = 2.4$ & $\alpha = 2.6$ \\
\hline $\nu_\mu$ & 0.27 & 0.23 & 0.21 & 0.18 \\
\hline $\anu_\mu$ & 0.26 & 0.23 & 0.20 & 0.18 \\
\hline $\nu_e$ & 0.27 & 0.24 & 0.22 & 0.19 \\
\hline $\anu_e$ & 0.26 & 0.23 & 0.20 & 0.18 \\
\hline $\nu_\tau$ & 0.27 & 0.23 & 0.21 & 0.18 \\
\hline $\anu_\tau$ & 0.26 & 0.23 & 0.20 & 0.18 \\
\hline
\end{tabular}
\end{center}
\caption{$\nu/\gamma$ ratios as seen at Earth after propagation.}
\label{tab:nug}
\end{table}
As an effect of oscillations the flavor composition is almost isotropized: the expected fluxes for each flavor are the same within a 20\%. Another important effect is the appearing  of a
  $\nu_\tau$ component, even if it  is not generated at the source.
  
 Neutrino telescope looking for up-going muons produced in the Earth are mainly sensitive
  to the incoming flux of $\nu_\mu$ and ${\bar \nu}_\mu$ .
 As a reference we write here the relation between the arrival $\nu_\mu +{\bar \nu}_\mu$  and $\gamma$-ray fluxes from a SNR accelerating primary protons with 
  a power-law spectrum with $\alpha = 2.2$
  \begin{equation}
 \label{flux_ratio}
 F_{\nu_\mu + {\bar \nu}_\mu}(E_\nu) \vert_{\alpha=2.2}
   = 0.46 F_{\gamma}(E_\nu)~.
 \end{equation}
  
We will not  discuss here the errors introduced by the uncertainties in our knowledge of the
 mixing angles.   As showed in \cite{Costantini:2004ap}, they are much smaller
 than the expected experimental errors.
 
It is important to notice that since ''averaged'' oscillations are energy independent
they  do not affect the power-law energy spectrum expected for neutrino at the source.
 As it has been pointed out in \cite{Crocker:2001zs}, a spectral distortion in the flux of astrophysical neutrinos might be the signal of long-wavelength energy-dependent oscillations, possibly due to the existence of new families of sterile neutrinos.
    
 \section{Antineutrinos from neutron decay}\label{antinu}
    
The interest about this antineutrino production mechanism has been stimulated by
claims of  possible detections of CR anisotropy in the EeV range.
Namely, the HIRES \cite{Hires} and AGASA \cite{Hayashida:1998qb} collaborations 
found  an excess of CRs with energy close to the EeV coming from the direction of 
 Cygnus-OB2, though with low statistical significance.
CR excesses was also detected by the AGASA \cite{Hayashida:1998qb} and SUGAR \cite{Bellido:2000tr} experiments in directions close to the Galactic Centre
(GC) and in the same energy range. Even if the existence of this anisotropy has not been confirmed by AUGER \cite{AUGERany}, theoretical arguments suggest that
a positive signal may be found when the statistics of that experiment
will be improved \cite{Grasso:2005wd}.
Both the Cyg-OB2 star association, laying in the Galactic plane at a distance of $1.7~\kpc$ from the Earth, and the region within 100 pc from the GC, at a distance of about $8.5~\kpc$,
are known to be active star forming regions rich of molecular gas and dust.
The origin of the CR anisotropy in their directions have been imputed to a neutron emission from those systems with energy up to few EeV. At that energy, in fact, neutrons can travel over galactic distances and, differently from protons and composite nuclei which 
are isotropised by the galactic magnetic field, they should keep the  information about 
the angular position of their source unspoiled. It was proposed that if primary nuclei are accelerated above the EeV in that system, their PD onto the Far Infra Red (FIR) radiation which is present in Cyg-OB2 and in the GC regions could efficiently produce
neutrons of the required energy\cite{Anchordoqui04}.
The presence of a FIR background  is a common feature of star forming regions.
It is due to the absorption and reprocessing of optical and UV stellar radiation 
by the thick dust which is generally present in those systems. The spectrum of this 
radiation is almost that of a black-body with temperatures typically ranging between 
10 and 100 K.

The authors of \cite{Anchordoqui04} suggested that, would the detection of a neutron flux
from Cyg-OB2 and the GC be confirmed, a ``guaranteed" flux of antineutrino in the TeV
energy range has also to be expected from those sources.
Their argument is based on the hypothesis that the primary nuclei spectrum
 between 1 PeV and few EeV is a rather steep power law (they assumed $\alpha \simgeq 3$
 as for the CR spectrum observed on the Earth)
 and that TeV antineutrinos are produced by the decay of neutrons, with energy close to the
 PeV, which are produced by the PD of nuclei onto the UV radiation in the proximity of massive stars.
 Although this is an interesting possibility, it should be noted that the neutron spectrum was
 not measured and that, since neutrons should be produced in the nearby of the source
and do not diffuse in the Galactic magnetic field, it is questionable that its slope coincides 
with that of ordinary CRs.
 Diffusion of primary nuclei in the MC region should be almost absent since, even assuming
 a magnetic field strength of $\sim 100~\muG$, their giration radius at energies above the 
 EeV is comparable to the cloud size.
 Furthermore, only few, if any, SNRs may be able to accelerate nuclei beyond the EeV.
  
 Here we consider a more general scenario in which neutrons may not be produced
 with the energy needed to reach the Earth and be detectable and  try to relate the antineutrino flux from neutron decay to the $\gamma$-ray flux 
which may be observed coming from the same source.
 We will determine the neutron production rate in the nearby of the SNR by applying some
 of the results derived in a previous work \cite{Grasso:2005wd}.
As secondary neutrons can be produced not only by the PD
of high energy nuclei but also by hadronic collisions, mainly
 $p + p_{\rm gas} \rightarrow n + p + {\rm pions}$, both processes need to be considered.
 We will show that, although PD is expected to  be the most effective production process of 
 EeV neutrons in MCs, at lower energies the hadronic scattering channel should be the 
 dominant neutron production channel.
  
\subsection{Neutrons from nuclei photo-disintegration}\label{PD}
  
 SNRs are expected to accelerate nuclei by a first order shock acceleration
 mechanism. In the linear shock acceleration theory all nuclear species are expected to
 have approximately the same power-law energy spectrum up to some UV cut-off depending on the charge number $Z$.

Above a centre of mass energy of a few MeV, a nucleus can photo-disintegrate
into a lighter nucleus and one or more nucleons, due to the resonant giant dipole interaction with
background photons. This process was studied in details in \cite{Puget:76,Stecker:99}.
Starting from the results presented in those papers, some of us  derived the neutron
 production rate due to the PD of nuclei onto a thermal background of photons at the temperature
 $T$ \cite{Grasso:2005wd}. This is
\begin{equation}
\label{PDrate}
{\Gamma}_{A} (E, T) \simeq 4 \times 10^{-14} ~ \xi_{A} \Phi_{A}(E, T)
~ \left(\frac{n_\gamma}{10^2~\cm^{-3}}\right) \; \s^{-1}~,
\end{equation}
where $\xi_A$ is an order one parameter depending on the nuclear species.
 The function $\Phi_{A}(E, T)$, which has been defined in  \cite{Grasso:2005wd}, 
  depends on the product $E T$ only. It is peaked  
  at the energy
\begin{equation}
\label{E_A_max}
E_A^{*} \simeq A~ 2\times
10^{15} ~ \left( \frac{T}{2\times 10^4 ~{\rm K}} \right) ^{-1}~\eV~
\end{equation}
where it takes a value depending on the nuclear species. For Iron nuclei ($^{56}$Fe)
$\Phi_{56}(E^{*}) \simeq 1$ , it is an order of magnitude smaller for
$^{16}$O and $^{12}$C, and $3 \times 10^{-2}$ for $^4$He.
Although $^4$He are the most abundant composite nuclei, heavier species, especially Iron, give a comparable contribution to the PeV neutron production yield due to their larger PD
cross section.
 
Equation (\ref{PDrate}) has been derived under the assumption that the photon
distribution is that of a black-body, or a grey-body (diluted black-body), with temperature $T$.
 Since each neutron takes, on average, a $1/A$ fraction of the parent nucleus energy,
 it follows from (\ref{E_A_max}) that PeV neutrons will be mainly produced
 by  nuclei PD onto a photon background with $T \sim 1~\eV$.
 However, due to the huge opacity of the dense molecular gas practically no optical and UV
photons should be present in GMC but in small H\,II blisters produced by the photoionization
of the hydrogen gas around massive stars. The radius of these regions, which are generally
known as Str\"omgren spheres, is expected to be roughly (see e.g. \cite{Shore})
\begin{equation}
R_{\rm H~II} \simeq 0.2 \left(\frac{F}{10^{48}~\s^{-1}}\right)^{1/3}
\left(\frac{\beta}{2.6 \times 10^{-23} ~\cm^{-3}\s^{-1}}\right)^{-1/3}
\left(\frac{n_{H}}{10^{3}~\cm^{-3}}\right)^{- 2/3}~\pc~,
\end{equation}
where $F$ is the UV photon production rate of OB stars and $\beta$ is the recombination
coefficient. The size of the Str\"omgren spheres grows relatively little during the short main
sequence life-time of these massive stars \cite{Shore}. Furthermore, the contribution of dust to the opacity may further reduce $R_{\rm H\,II}$.
In dense GMCs, having $n_H = 10^3 - 10^5 ~\cm^{-3}$ a characteristic size of
$10~pc$ and a population of hundreds of OB stars, the H~II phase filling factor
is smaller than $10^{-2}$. Since the photon density in each blister is
$\displaystyle n_{\rm UV}^{\rm H~II} \simeq \frac{F}{\pi c R_{\rm H\,II}^2}
\simeq 10^2~\cm^{-3}$, the overall mean UV photon density in a dense MC is
roughly $ n_{\rm UV}^{\rm MC}\simeq 1~\cm^{-3}$.
An order of magnitude larger photon density may be found in rich systems like the
Cyg-OB2 association.
As a consequence, in such a kind of systems, the H\,II phase filling factor is close to unity
and the UV photon density $n_{\rm UV} \sim 10~\cm^{-3}$ which
  is in agreement with the value adopted in \cite{Anchordoqui04}.
Using (\ref{q_gamma}) and (\ref {PDrate}) we find
\begin{equation}
\label{Q_n_PD}
\begin{split}
& Q_n(E_n) = \sum_A f_A ~ n_p(E_n) ~ \Gamma_A(A E_n, T) \\
 & = 4\times 10^{-17} \left(\frac{\sum_A f_A \Phi_A(E_n, T)}{10^{-2}}\right)
 ~n_p(E_n) ~\left(\frac{n_\gamma}{10~\cm^{-3}}\right)~\GeV^{-1} \cm^{-3} s^{-1}
\end{split}
\end{equation}
where $f_A$ is the relative abundance of all the nuclear species with atomic weight $A$
in the SNR which undergo effective PD. For the proton spectrum we used the unit 
 $ \GeV^{-1} \cm^{-3}$.
 
 \subsection{Neutrons from p-p scattering}\label{neutroni_pp}

Besides the PD process discussed in the previous section, EeV neutrons
might also be produced by the collision of UHE nuclei with the dense gas in the MC
environment. The relevant neutron production channel is the charge exchange $pp$ inelastic
collision $pp_{\rm gas} \rightarrow p n + {\rm m}_\pi \pi$, where $ {\rm m}_\pi$ is the pion multiplicity.
In the case of a power-law spectrum for the HE protons
the neutron emissivity due to this process was found to be \cite{Grasso:2005wd}
\begin{equation}
 \label{Qn_pp}
\begin{split}
Q_n^{pp}(E_n) & = n_H n_p(E_n) \sigma_0 ~{\rm m}_n~ c~ K~ I(\alpha)\\
& \simeq 2 \times 10^{-15} ~n_p(E_n)~\left(\frac{n_H}{10^3~\cm^{-3}}\right)
\left(\frac{I(\alpha)}{0.1}\right)~\GeV^{-1} \cm^{-3} s^{-1}
\end{split}
\end{equation}
where $K =4/5$ is the inelasticity, and ${\rm m}_n \simeq 0.24$  the neutron multiplicity.
We neglected here a weak logarithmic energy dependence of the cross section.
The function $I(\alpha)$ is defined by
\begin{equation}
\label{Ialpha}
I(\alpha) \equiv \int_0^1dx\,x^{\alpha-2}h(x)
 \end{equation}
where $h(x) \simeq 0.064\left(1-x\right)^2+0.094x\sqrt{1-x}$ and $x \equiv E_n/E_p$.

By comparing (\ref{Qn_pp}) with (\ref{Q_n_PD}) it is evident that under most common
conditions in GMCs, $pp$ scattering should be the main neutron production channel at
energies around the PeV.

An approximated expression for the antineutrino emissivity from neutron decay can then be derived by treating the neutron decay factor as a step function $\displaystyle
 \theta \left( \frac{ d m_n}{\tau_n} - E_n \right)$ and replacing the antinuetrino
 energy distribution by a delta-function peaked at the mean value ${\bar E}_{\anu} \simeq
   6 \times 10^{-4}~m_n$ (here we followed the same approach adopted in \cite{Anchordoqui04} ).
 By doing that, we find
  \begin{equation}
\label{F_anu_approx}
Q_{\anu} (E_{\anu}, {\bf x})
= 0.1 ~\frac{I(\alpha)}{Y(\alpha)}~
\left( \frac{m_n}{2 {\bar E}_{\anu} }\right)^{1 -\alpha}~Q_\gamma(E_{\anu}, {\bf x})\; .
\end{equation}
Comparing this result with that given in Sec.\ref{sec:nuPion}, it is clear that, unless the primary
 proton spectrum has a quite unusual behaviour, neutron decay should give only a negligible
 contribution to the total neutrino emissivity of an embedded SNR.
 
 The relative contribution of neutron decay might somewhat be enhanced if the proton energy distribution in the MC has a strong spatial dependence.
 This may be indeed the case for protons which are shock accelerated by a steady SNR, as
 the expected energy spectrum upstream the shock front is \cite{Malkov}
\begin{equation}
\label{malkov}
n_p(E,r) = n_p(E,0)~\exp\left\{ - \frac {\kappa~ u_S~r}{D(E)}\right\}~.
\end{equation}
 Here $r$ is the distance from the SNR shock;
 $n_0(E) $ is the proton spectrum downstream the shock, which we assume to be given by
  (\ref{p_spectrum}) ; $u_S$ is the shock velocity;
 $D(E)$ is the diffusion coefficient; $\kappa$ is an order one coefficient.
 For the sake of simplicity we also assumed that $u_S$ and $D(E)$ do not depend on $r$.
By expressing the energy dependence of the diffusion coefficient in the usual form
$D(E) = D(E_0) (E/E_0)^\beta$, the meaning of (\ref{malkov}) become quite evident:
upstream the shock the proton spectrum has an exponential low-energy cutoff at
\begin{equation}
E_{\rm min}(r) = E_0 \left( \frac{ u_S ~r}{D(E_0)} \right)^{1/\beta}~.
\end{equation}
Assuming Bohm diffusion in the acceleration region (note that adopting a larger diffusion
coefficient in that region would imply a too low maximal acceleration energy),
  i.e. $\displaystyle D(E) = \frac{1}{3}~c~r_L(E) $, where $r_L(E)$ is the Larmor radius, we find
\begin{equation}
 r_*(E) \simeq 20 \left(\frac{E}{1~\PeV}\right)
\left( \frac{u_S}{500~{\rm km}~\s^{-1}}\right)^{-1}
\left(\frac{B}{10~\muG}\right)^{-1} ~\pc~.
\end{equation}
  It is evident that PeV protons can diffuse into the cloud much more deeply than $10~\TeV$, so
that secondary neutrons,  as well as TeV $\anu_e$ from their decay,  are produced in
a volume larger than that in which photons and neutrinos are emitted by hadronic scattering.
  Note, however, that the characteristic distance over which protons propagate before loosing
their energy due to $pp$ scattering ($\tau_{\rm loss} \simeq (c~n_H \sigma_0 )^{-1}$), is
\begin{equation}
L_{\rm loss}(E) = \sqrt{6 D(E) \tau_{\rm loss}} \simeq 10~ \left(\frac{E}{1~\PeV}\right) ^{1/2} \left(\frac{B}{10~\muG}\right)^{-1}
\left(\frac{n_H}{10^{3}~\cm^{-3}}\right)^{-1/2}~\pc~.
\end{equation}
Therefore, energy losses significantly  reduce the effective propagation volume of primary 
  protons in those cases in which the gas density and magnetic field strength are very large.
 Unfortunately, these are also the most interesting cases from the observational point of view.
At the end, we found that in all relevant physical situations the volume gain factor is not sufficient
to make the enhancement of the contribution of neutron decay to the total neutrino emissivity
observationally relevant.
 
 We conclude this section by stressing that all the previous considerations apply only if the 
 particle primary spectrum is  a  power law.
 This will not  be the case, for example, if primary nuclei are accelerated
by a strong voltage drop around a fast rotating neutron star. 
  In that case, in fact, primary nuclei  may  have an energy  spectrum  peaked at the PeV, 
  or above,  so that their PD onto stellar radiation be dominant respect to  hadronic scattering and   neutron decay  be the main neutrino production channel.  
  
\section{Neutrinos from the Galactic Centre}\label{sec:nuGC}

In this section we apply our previous results to estimate the neutrino flux from the
dense molecular cloud complex in the Galactic Centre region.
An extended very-high-energy $\gamma$-ray emission has recently been discovered by the HESS
coming from a region roughly delimited by $|l| < 0.8^{\circ}$ and $|b| < 0.3^{\circ}$ in galactic coordinates \cite{Hess2006}.
The coincidence between the angular position of the centroid of this emission and that of the
dynamical centre of the Galaxy suggests that both are located at the same distance from the
Earth, which is about $\sim 8.5~\kpc$.
The energy spectrum of this source was measured to be 
$(1.73 \pm 0.13 \pm 0.35)\times \displaystyle
E_{\TeV}^{- (2.29 \pm 0.07 \pm 0.02)}~\TeV^{-1} \cm^{-2} \s^{-1} {\rm sr}^{-1}$.
Superimposed to that diffused emission HESS found a point-like source (J1745-290), which
 was also early detected by HESS \cite{Hess2004} (and confirmed by MAGIC 
\cite{Albert:2005kh}) with an energy spectrum $\displaystyle
  (2.50 \pm 0.2) \times E_{\TeV}^{- (2.21 \pm 0.09)}~\TeV^{-1} \cm^{-2} \s^{-1}$,
 in a position compatible either with the supermassive black hole in Sgr A$^*$ or with the SNR
  Sgr A East. This source was subtracted when determining the spectrum of the extended emission.

As the authors of \cite{Hess2006} suggest, the detected $\gamma$-rays are likely to be produced
by the hadronic interaction of locally produced cosmic rays with the dense molecular gas.
A relevant support in favour of this hypothesis is given by the strong correlation found between the
$\gamma$-ray and the millimetric   emissions from the CS (Carbon Sulfide) 
 in that region, that compound being a well known tracer of molecular hydrogen.
The coincidence, within errors, between the spectral slope of the GC ridge emission
 detected by HESS and that of J1745-290, suggests that the primary particles responsible
 for both signals may have the same origin.
 Indeed, the total CR energy required to explain both of them is compatible with that
typically expected from shock acceleration by the remnant of a single SN.
These considerations point to Sgr A East as the most plausible primary source
(see \cite{Crocker:2004bb, Grasso:2005wd} about the identification of J1745-290 with that SNR).
 
This elegant scenario may find further support would neutrinos from the same region
be detected.
The expected neutrino flux can be easily determined by applying the results of Sec. \ref{sec:nuPion}.
It has to be noted that due the relatively poor angular resolution (of order $\simleq 1^{\circ}$)
achievable with Neutrino Telescopes and the very low expected neutrino fluxes, it will be quite
hard, if not impossible, to disentangle the neutrino point-like emission from that coming from the
 Galactic Centre ridge.
Thus, we have to sum the neutrino flux from both sources. This is straightforward since, as the $\gamma$-ray spectral slopes of these sources coincide within errors, the same will
be true for neutrinos. We use here the central value of the slope measured by HESS for the
GC ridge, i.e. $\alpha = 2.29$.
The total $\gamma$-ray flux from the GC ridge source is determined by multiplying the diffused
flux measured by HESS by its solid angle opening, i.e. $4 \times 0.8^{\circ} \times 0.3^{\circ}
= 3 \times 10^{-4}~{\rm sr}$. It is about the double of the flux from J1745-290.
Therefore, the total $\gamma$-ray flux, above the TeV, coming from the GC region is roughly
(we only write central values in the following)
\begin{equation}
F_\gamma^{\rm GC}(E_\gamma) \simeq 7.5 \times 10^{-12}
\left(\frac{E_\gamma}{1~\TeV}\right)^{-2.3} \,\TeV^{-1}\,\cm^{-2}\,\s^{-1}~.
\end{equation}
 Using our results of Sec.\ref{sec:nuPion}, we derive the $\nu/\gamma$ ratios for
 $\alpha = 2.29$ at the source and at the Earth are shown in Tab.\ref{tab:rat} for the
 different neutrino species. We notice here that the TeV $\gamma$-ray attenuation due to pair-production scattering onto the IR photon background should be negligible. In fact, whenever the radiation temperature is below 100 K, the mean IR photon energy implies a $\gamma$-ray threshold energy well above 10 TeV. Gamma-ray scattering onto the UV and X-ray backgrounds present in that region was also showed to give rise to a negligible attenuation \cite{Crocker:2004bb}.
\begin{table}
\label{tab:rat}
\begin{center}
\begin{tabular}{|c|c|c|}
\hline $\alpha = 2.29$ & at source & at Earth\\
\hline$\nu_e$ & 0.24 & 0.23\\
\hline$\anu_e$ & 0.21 & 0.21\\
\hline$\nu_\mu$ & 0.43 & 0.22\\
\hline$\anu_\mu$ & 0.43 & 0.22\\
\hline$\nu_\tau$ & 0 & 0.22 \\
\hline$\anu_\tau$ & 0 & 0.22 \\
\hline
\end{tabular}
\end{center}
\caption{$\nu/\gamma$ ratios before and after oscillations.}
\end{table}

It is now straightforward to derive the total expected muon neutrino flux which is
\begin{equation}
\label{flux_nuGC}
F_{\nu_\mu+\anu_\mu}^{\rm GC}(E_\nu) \simeq 3.3\times 10^{-12}
\left(\frac{E_\nu}{\TeV}\right)^{-2.3}\,\TeV^{-1}\,\cm^{-2}\,\s^{-1}~.
\end{equation}
This flux is about the double of that estimated in \cite{Crocker:2004nk} using only
the 2004 HESS data \cite{Hess2004}.
Since the detection of secondary up-going muons produced by neutrino interaction in the Earth is
the most promising strategy to identify TeV neutrino from astrophysical sources, Neutrino Telescopes in the North hemisphere are best suited to look for a signal from the GC as it is
located in the southern Sagittarius constellation.
 Using the $\nu_\mu +\anu_\mu$ flux given in (\ref{flux_nuGC}), 
 and the ANTARES effective area for up-going $\nu_\mu$s   given in \cite{ANTARES}, 
 we estimated the expected
 detection rate by this experiment to be  $0.07~{\rm yr}^{-1}$ which is, unfortunately, 
 quite small. A ${\rm km}^3$ Neutrino Telescope  to be built in the Mediterranean sea (e.g. NEMO \cite{NEMO} or NESTOR \cite{NESTOR} upgrading) has better chance to 
 get a positive detection. Indeed,  since the  effective area of such an instrument should be
 at least 20 times better than that of ANTARES,  the  expected rate is  $1.5 ~{\rm yr}^{-1}$.   
As  the expected background rate of such device is $\sim 0.3~{\rm ev/yr}$ . \cite{Crocker:2004nk}, the observation time required
to detect that source  at 95\%C.L is about three years.

\section{Conclusions}

In this paper we estimated the neutrino/photon flux ratio in the TeV energy range from {\it
embedded SNRs}. We did that under the hypothesis that SNRs shock accelerate protons and
composite nuclei with a power law energy spectrum at least up to the PeV. By assuming that all
$\gamma$-rays of that energy are produced by $\pi^0$ decay, we first computed the relative
 emissivities of muon and electron neutrinos produced by charged pion decay.
Then, we took into account for vacuum oscillations to determine the $\nu_i/\gamma$ flux ratio
to be expected on the Earth for all lepton families. Our results are
similar, though not coincident, to those previously derived by other authors \cite{Costantini:2004ap,Crocker:2004nk}.
We also considered the possible contribution that neutron decay may give to the TeV antineutrino
flux from a GMC embedding an active SNR. Using the results which we derived in a previous work
\cite{Grasso:2005wd}, we showed that in dense molecular complexes (e.g. in the GC)
 hadronic scattering should be the dominant neutron production mechanism at the relevant energies ($\sim \PeV$). That channel was not considered in \cite{Anchordoqui04}.
Unfortunately, we found that, although such a neutrino production channel is less severely constrained by $\gamma$-ray observations, it is quite unlikely that under reasonable assumptions it could add a significant contribution to the total neutrino flux from such kind of
sources.

As an application of our results we used the recently released data of the GC observation campaign
by the HESS collaboration \cite{Hess2006} to estimate the muon neutrino flux from that region.
We found that the ${\rm km}^3$ Neutrino Telescope to be built in Mediterranean sea
may have some chances to detect that source.  Although the expected
neutrino detection rate is rather small, the possible observation of a positional coincidence
between the $\gamma$-ray and the neutrino source as well as the measurement of a flux ratio
 at the level estimated in this paper, would add significant evidence in favour of the hypothesis
 that hadron acceleration is taking place in that region.
  Furthermore, since embedded SNRs should be quite common objects in the Galaxy and
 VHE $\gamma$-ray and neutrino astronomy have just  started, 
 other sources of that kind may be discovered in both channels in a not too far future. 
  
\section*{Acknowledgements}
 
 We  thank V. Flaminio, T. Montaruli, T. Sjostrand and F. Vissani for several valuable discussions. 
 The authors are members of the ANTARES collaborations.   L.M. thanks G. Usai for help using the \textsc{Pythia}  package.

 \end{document}